# Preparation of high-precision aspherical lenses based on micro stereolithography technology

Lu xiaoying ,Liu hua

School of Physics, Northeast Normal University, Changchun, Jilin Province

**Abstract** The 3D printing technology based on digital light processing (DLP) has highlighted its powerful manufacturing capabilities for optical components. However, the printing structure obtained by DLP based down projection printing is easily adhered to the printing window below, and the printed lens surface will have a step effect. This article uses DLP 3D printing technology to print non spherical lenses. During the printing process, a new type of inert liquid fluoride solution was used as the isolation layer, which can more effectively and conveniently prevent the printing structure from sticking to the printing window. At the same time, a vertical lifting immersion method was proposed to smooth the step effect on the surface of the lens.

## 0Preface

Compared to traditional optical components, non spherical optical components have more design degrees of freedom, which can effectively improve the performance of optical systems, simplify the structure of optical systems, and be applied in multiple fields such as automobiles [1,2], medical [3-5], lighting, etc. The traditional manufacturing method for aspherical optical components has a long processing cycle, which limits the widespread application of aspherical optical components. 3D printing technology, also known as additive manufacturing technology [6-9], is one of the rapidly developing green manufacturing technologies in recent years. The 3D printing process can significantly reduce the cost of prototype design, process complex structures that traditional processing techniques cannot manufacture, and reduce material waste [10]. Among various 3D printing technologies, the high precision and fast speed DLP 3D printing technology has highlighted its powerful optical component manufacturing capabilities [11-17].

However, there are also certain limitations to the DLP 3D down projection printing technology. The most obvious one is that DLP 3D printing creates physical objects by stacking two-dimensional layers of slices together, which creates a step effect on the resulting surface [18-20]. In order to eliminate the step effect, Kang [21] used a method of combining equal arc and equal thickness slicing with continuous printing to smooth the surface smoothness and successfully prepared millimeter level spherical lenses. However, the slicing method is somewhat complex and requires flexible switching based on the surface steepness of specific components. The most common step processing method currently is the meniscus method proposed by Pan et al. [22]. Modeling can be achieved by considering the effects of capillary action, gravitational attraction, multi-layer adsorption, and boundary conditions [23,24], and matching the meniscus of each layer with the predetermined modeling surface, thereby reducing the impact of step and pixel effects on printing results [25]. However, the liquid film formed by this method involves multiple control parameters, and the thickness of the liquid film is difficult to control. At the same time, for relatively viscous liquids, it takes a long time to level. For small-sized lenses placed on larger flat substrates, due to the boundary constraints of the substrate on the liquid, ideal liquid thin films cannot be formed. There is also the use of grayscale methods to eliminate step effects. Xu [26] used LCD screens to generate digital grayscale masks, and combined them with the meniscus method to achieve smooth optical surfaces. Chen et al. integrated grayscale photopolymerization technology and meniscus balance post curing technology to print non spherical surfaces with surface roughness below 7nm. However, grayscale processing requires high demands on photosensitive materials and the relationship between curing thickness and exposure energy is non-linear, making it difficult to accurately control the relationship between grayscale and curing thickness.

In this article, we propose a simpler and more efficient method to eliminate the step effect on the surface of the lens, which is achieved through the vertical lifting immersion method. Explored the relationship between pulling speed and film thickness, and finally used a speed of 14.6mm/s to pull, which can achieve critical resin filling "steps" and effectively improve the imaging quality of the lens. Successfully manufactured customized aspherical lenses with a height of 2mm and a diameter of 5mm. The optical performance was characterized through experiments, and the application pathways of the prepared lens were supplemented. In addition, this article uses fluoride solution as a constraint interface to eliminate printing structure and bottom adhesion, improving printing efficiency and effectiveness.

# 1. DLP 3D Printing System Construction

## 1.1 Experimental equipment and workflow

The DLP system built by the laboratory is shown in Figure 1a, which mainly consists of a computer, a UV laser light source (386 nm), a green LED lighting source (for focusing), a double lens from Edmond Optics, a digital micro mirror device (DMD, Texas Instruments) from Texas Instruments, a CCD camera (Edmund EO camera), a motion axis, and piezoelectric components.

Taking the downward projection DLP printing system used in the laboratory as an example, introduce the workflow of the system. Firstly, place an appropriate amount of fluoride solution into the resin tank. The fluoride solution is transparent, has a higher density than the resin, is highly immiscible with the resin, and has low surface energy, which can reduce the mechanical separation force as shown in Figure 2 (a). Due to the uniform dissolution of oxygen in the fluoride solution, a flat dead zone interface is formed, which is an unreacted area formed by the oxygen inhibition of the fluoride solution. When fresh resin flows into the polymerization zone composed of dead zone and curing zone, the fluorination solution can serve as an interface for suppressing photopolymerization and dehumidification, improving the surface smoothness of the 3D printing structure, as shown in Figure 2 (b) (c). Afterwards, we shifted the focus to the surface of the fluoride solution and filled the resin tank with a large amount of prepared resin material. Transfer the printed model to the computer, which slices the model and then transfers the layered image to the projection device. The projection device will control ultraviolet light based on the mask image to image the layered image on the lower surface of the photosensitive resin liquid. Firstly, reset the printing platform and project the first mask image. The photosensitive resin near the surface of the fluoride solution undergoes a photopolymerization reaction after being exposed to ultraviolet light. At this point, the single-layer molding work is completed. Afterwards, the motion axis Z began to lift under the control of the piezoelectric platform. We can adjust the speed and step distance of the Z-axis on the computer. The so-called step distance refers to the distance set for each Z-axis movement, and the Z-axis will transmit a signal to the DMD. When the DMD receives the signal, it will immediately replace the next mask image, as shown in Figure 1b.

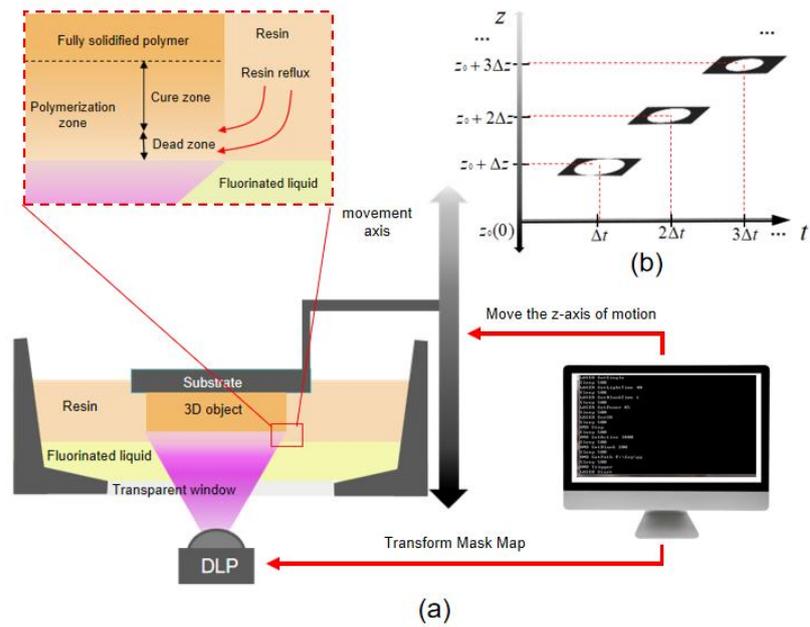

Figure 1 (a) DLP system (b) DLP 3D printer mechanism explanation

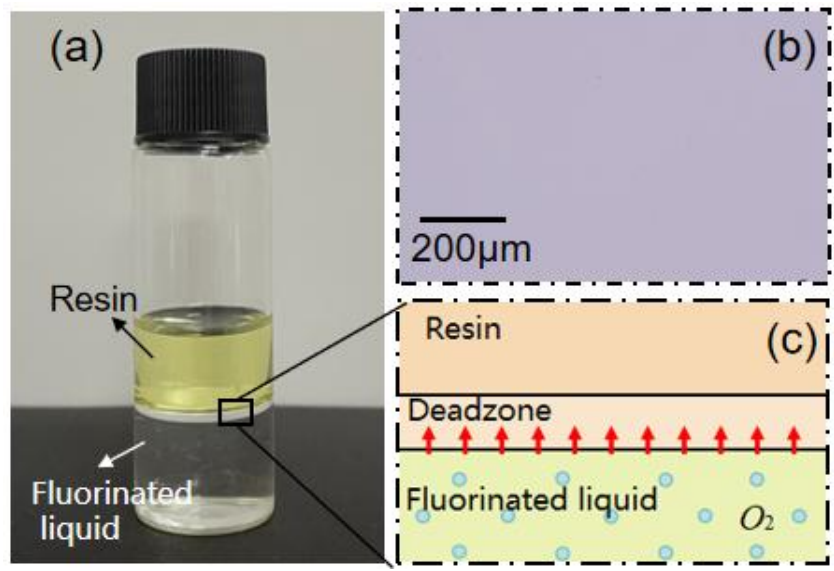

In Figure 2 (a) of the glass tube, the upper liquid is resin and the lower liquid is fluorinated liquid. (b) The smooth surface of the fluorinated liquid seen under a 10 x microscope. (c) Oxygen is uniformly dissolved in the fluorinated liquid, forming a smooth dead zone interface

## 2 Exploring Parameters and Smoothing Methods Based on DLP 3D Printing Technology

### 2.1 Configuration of Printing Materials

The resin monomer is the main body that undergoes photopolymerization, so the selection of monomers is crucial. The resin monomer used in this study is 1,6-hexanediol diacrylate (HDDA), which has two acrylic groups that can participate in the reaction [28-30]. It has the same functionality, fast curing speed, low viscosity, and good solubility. Considering the speed of the

UV curing process, a photoinitiator was added to the monomer. The photoinitiator chosen for the experiment was phenyl bis (2,4,6-trimethylbenzoyl) phosphine oxide (819, C26H27O3P). To reduce the curing depth, a UV absorber was added to the monomer, which is a light stabilizer that can absorb the ultraviolet part of sunlight and fluorescent light sources without undergoing any changes. The incident ultraviolet light will be greatly attenuated by the ultraviolet light absorber, and less light can penetrate deeper to cause photopolymerization. The absorbent selected for the experiment is 2- (2H benzotriazol-2-yl) -6-dodecyl-4-cresol (171, C25H35N3O).

In order to investigate the effects of photoinitiators and absorbers on the size and single layer curing depth in the X and Y directions, two layers of circles with a mask image diameter of 2500 μm and 2050μm were exposed at the same time (single layer exposure time of 9 seconds) and energy (energy power of 4.4kw).

The influence of photoinitiators on the radius of a circle was investigated while ensuring the constant content of light absorbers. As shown in Figure 3a, with the increase of photoinitiator content, the size of the circle will also increase. During the experiment, it was found that when the initiator content exceeded 1%, the increase in initiator content did not have a significant impact on the radius of the circle. Finally, the content of the initiator was determined to be 1wt%. Under the condition of ensuring the constant content of photoinitiator, the influence of photoabsorbent on the curing depth was investigated. From Figure 3b, it can be observed that as the content of light absorbent increases, the curing depth decreases. In order to obtain a smoother lens surface in the later stage and reduce the step effect, an absorbent content of 5wt% was selected. Figure 3c shows the structure printed when the initiator content is 1% and the light absorbent is 5%.

## 2.2 Determination of printing parameters

Measure the printing structure using a microscope and calculate the relative size errors in the length (X), width (Y), and height (Z) directions to characterize the dimensional accuracy of the printing structure. The printing structure is a rectangular prism with a length of 2394 μm, a width of 1026 μm, and a height of 2000 μm. For each printed sample, measure the dimensional accuracy in each direction 5 times and calculate the average value as the actual value of the test sample. Relative error of size=| actual value - theoretical value |/theoretical value x 100%, which is the percentage of absolute error to theoretical value.

### 2.2.1 The influence of layer thickness on the dimensional accuracy of printed parts

Figure 3d shows the relative size error for different slice thicknesses. From the graph, it can be seen that as the layer thickness increases, the relative errors of the dimensions in the X, Y, and Z directions of the printed sample all show a trend of increasing from large to small and then gradually increasing. Moreover, the absolute error values of the dimensional accuracy in all directions change from positive to negative. The curing process of photosensitive resin follows Beer Lambert's law [25], as shown in equation (1).

$$C_d = D_p \ln \frac{E}{E_c} \quad (1)$$

In the formula, $C_d$ represents the curing depth; $D_p$ is the depth of light propagation; E is the exposure energy; $E_c$ is the critical exposure amount.

This is because as the layer thickness increases, the energy required for complete solidification of the single-layer liquid resin also increases. Under the condition of keeping other parameters constant, when the layer thickness is less than 8 μm, the thickness of the single-layer liquid resin is less than the curing depth Cd, and the liquid photosensitive resin is over cured, resulting in its actual size in the Z direction being larger than the design size. At the same time, excessive exposure energy E will solidify some of the horizontal liquid resin, causing the actual size in the X and Y directions to be larger than the design size; When the layer thickness is greater than 8 μm, the thickness of the single-layer liquid resin is greater than the curing depth Cd, and the resin is under cured, resulting in its actual size in the Z direction being smaller than the design size. At the same time, the exposure energy E of the horizontal edge contour part is not enough to cure the liquid resin, causing the actual size in the X and Y directions to be smaller than the design size. Finally, the layer thickness was determined to be 8 μm.

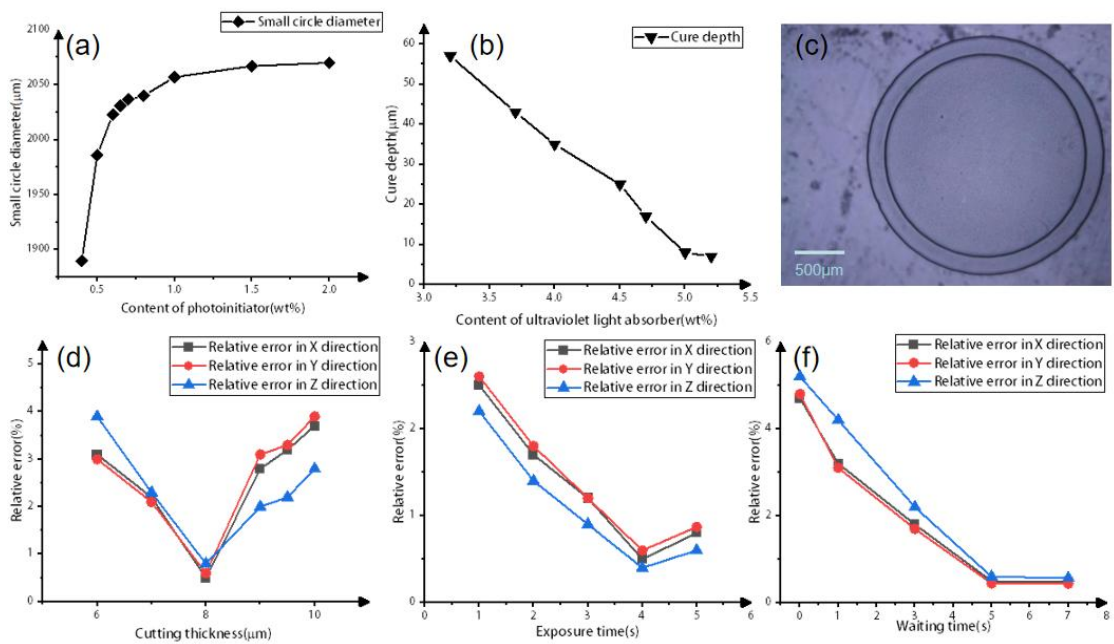

Figure 3 (a) Small circle diameter size obtained at different photoinitiator contents (b) Single layer curing depth obtained at different photoabsorbent contents (c) Printed structure at 1% photoinitiator content and 5% photoabsorbent content (d) Relative error at different slice thicknesses (e) Relative error at different exposure times (f) Relative error at different waiting times

**2.2.2 The impact of exposure time on the dimensional accuracy of printed materials**

Figure 3e shows the relative size error at different exposure times. From the graph, it can be seen that as the exposure time increases, the relative errors of the dimensions in the X, Y, and Z directions of the printed sample all show a trend of increasing from large to small and then gradually increasing. In addition, the absolute error values of the dimensional accuracy in all directions are from negative to positive. The exposure energy formula is given in equation (2).

$$E = It \qquad (2)$$

In the formula: I is the transmitted light intensity; T is the exposure time.

On the premise that the transmitted light intensity I remains constant, the longer the exposure time t, the greater the exposure energy E. Combining Beer Lambert's law and exposure energy formula, under the condition of keeping other parameters constant, when the exposure time is less than 4 seconds, the exposure energy E is not enough to make the curing degree of the single-layer liquid photosensitive resin reach the curing depth, resulting in the Z-direction dimension being smaller than the design value. At the same time, the exposure energy E of the horizontal edge contour part is not enough to cure the liquid resin, making the actual dimensions in the X and Y directions smaller than the design value; When the exposure time is 4 seconds, the liquid photosensitive resin is over cured, causing its actual size in the Z direction after curing to be greater than the design value. The excess exposure energy will cause the liquid resin in the horizontal direction to cure, causing the actual size in the X and Y directions to be greater than the design value. Finally, the single-layer exposure time was determined to be 4 seconds.

**2.2.3 The impact of single-layer waiting time on the dimensional accuracy of printed parts**

Figure 3f shows the relative error of printing size at different waiting times. From the graph, it can be seen that as the waiting time increases, the relative errors in the dimensions of the printing structure X, Y, and Z all decrease rapidly and tend to stabilize at a waiting time of 5 seconds. This is because the UV curing process converts liquid photosensitive resin into a solid state, and the change in molecular gap size leads to volume shrinkage and residual stress problems inside the workpiece. The setting of waiting time is conducive to the completion of volume shrinkage, reducing residual stress inside the workpiece, and ensuring that the size and shape of the workpiece will not undergo significant changes for a period of time after printing is completed. If the waiting time is too short, the resin molecules that have just cured in the previous layer have not fully completed the shrinkage process before proceeding to the next layer of curing, causing the actual size of the test sample to be larger than the design value, and the residual stress generated inside the sample will cause the sample to warp and deform; As the waiting time increases to a certain value, each layer of resin molecules can complete shrinkage, and the dimensional accuracy of the workpiece tends to stabilize. However, if the waiting time is too long, the efficiency of curing and forming is lower. Finally, the single-layer waiting time was determined to be 5 seconds.

**2.3 Use vertical lifting method to eliminate steps**

Like other additive manufacturing processes, objects manufactured by DLP technology also have a step effect. This is because we stack the two-dimensional layers of slices together to create physical objects that approximate the original model. Due to the stacking of the two-dimensional layers, the resulting surface may have significant errors. The vertical lifting immersion method can change the thickness of the liquid film attached to the surface of the lens by controlling the lifting speed.

During the printing process, the cured structure is constrained by the liquid resin interface, resulting in a thin liquid resin film wrapping around the outside of the printed structure. By appropriately controlling the restricted liquid residue film, it can fill the gaps between adjacent layers, thereby eliminating the step effect and obtaining a 3D structure with smooth sidewalls. Therefore, the immersion method is adopted to eliminate the step effect. We first cure a layer of polymer on the glass, then fix it on the Z-axis, and control the lifting speed of the Z-axis through a computer. This way, a thin film will be hung on the cured structure, as shown in Figure 4a. Afterwards, read the film thickness using a 10 x lens microscope. By changing the pulling speed, images of the thickness of the liquid film hanging at different pulling speeds were obtained, as shown in Figure 4b. The use in the application was obtained through fitting, which prevented the obtained lens from achieving good imaging results.

$$d = 0.01878 + 0.89649v - 0.02397v^2 \quad (3)$$

It can be observed that within a certain range of pulling speed, the thickness of the film on the surface of the cured resin increases with the increase of pulling speed. This is because the faster the speed, the higher the viscosity, the greater the upward viscous force, and the greater the downward gravity borne by the resin, resulting in a thicker deposited film [26]. In this experiment, the thickness of each layer of the step is 8 μm, so that d=8 μm. By substituting it into the fitting equation, the pulling speed at this time should be 14.6mm/s.

Fix the printed step structure on the Z-axis of motion and pull it vertically out of the resin, as shown in Figure 4c. Figures 4d and 4e show the cases of insufficient resin filling and critical resin filling when the speed is small and appropriate, respectively.

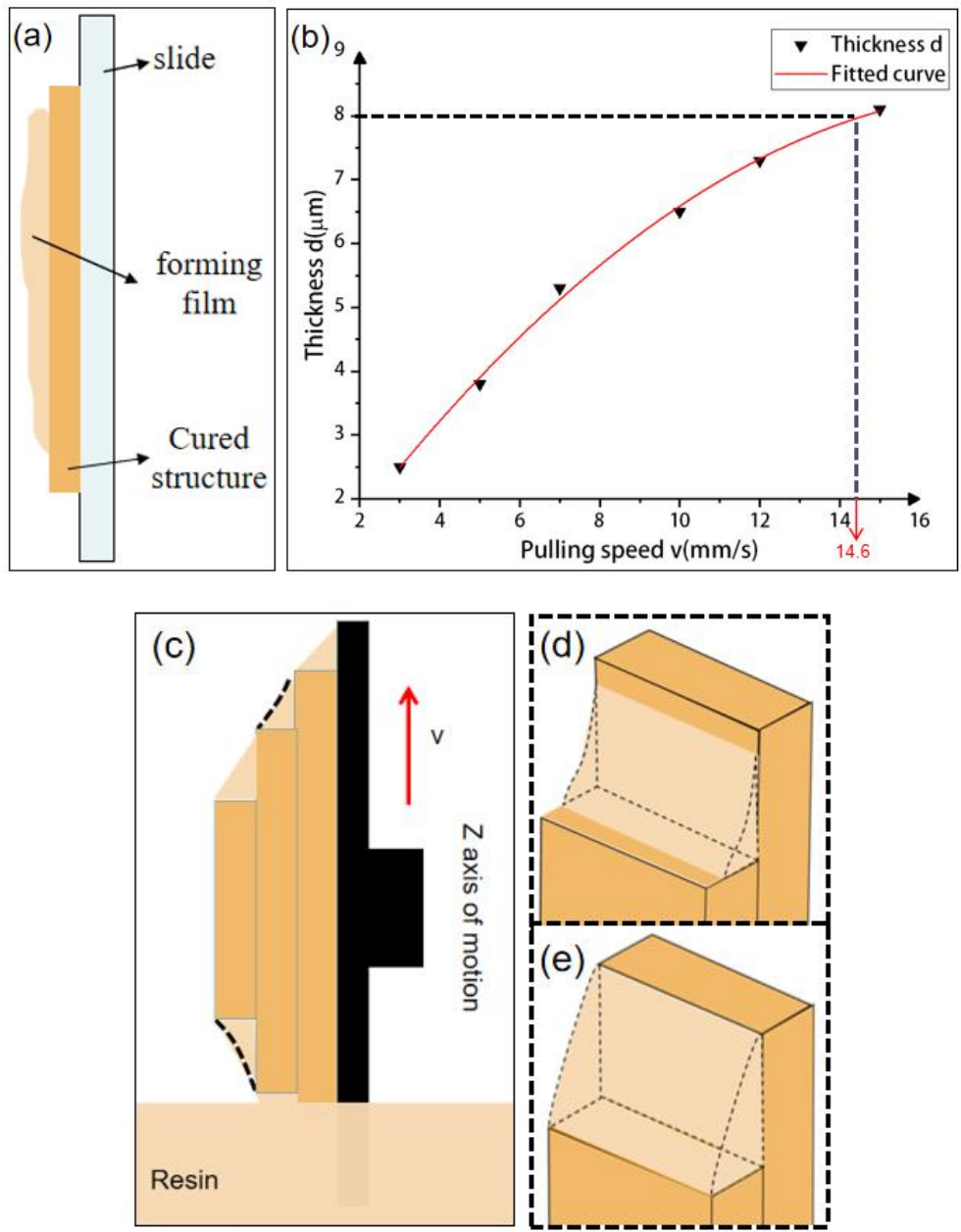

Figure 4 (a) Film hanging process diagram (b) Relationship between different pulling speeds and corresponding film thickness (c) Lens pulled out of resin using immersion method (d) Meniscus formed when resin filling is insufficient (e) Meniscus formed when resin filling is critical

## 3 Design, preparation, and testing of aspherical lenses

## 3.1 Design

We designed a simple lens using the optical design software ZEMAX. The aperture of the lens is 5mm and the thickness is 2mm. The imaging effect is shown in Figure 6a when the object distance is 20mm and the image distance is 22mm.

The surface equation of the design is represented as follows [31]

$$z(r) = \frac{cr^2}{1+\sqrt{1-(1+k)c^2r^2}} + a_1r^2 + a_2r^4 + a_3r^6 + a_4r^8 \tag{4}$$

Among them, c is the curvature (reciprocal of curvature radius), k is the cone constant, and r is the radial distance measured from the optical axis. The convex surface parameter designed is c=0.2, k=0，a1=0，a2=1.45E-003。

## 3.2 Preparation and Testing

According to the optimal printing parameters obtained, the stepped lens surface is printed, as shown in Figure 5a. After eliminating the steps using the immersion method (as shown in Figure 5b), the smooth lens obtained after secondary curing is shown in Figure 5c. The lens has a light yellow color tone, which is due to the UV absorption characteristics of the UV cured resin and can be further reduced as needed. In order to demonstrate the successful preparation of the printed lens structure and the elimination of step effects, the post-processed lens was removed and placed on a black and white grid, as shown in Figure 5d. When viewed through the lens, there was no distortion in the image, indicating its high smoothness and practical application potential.

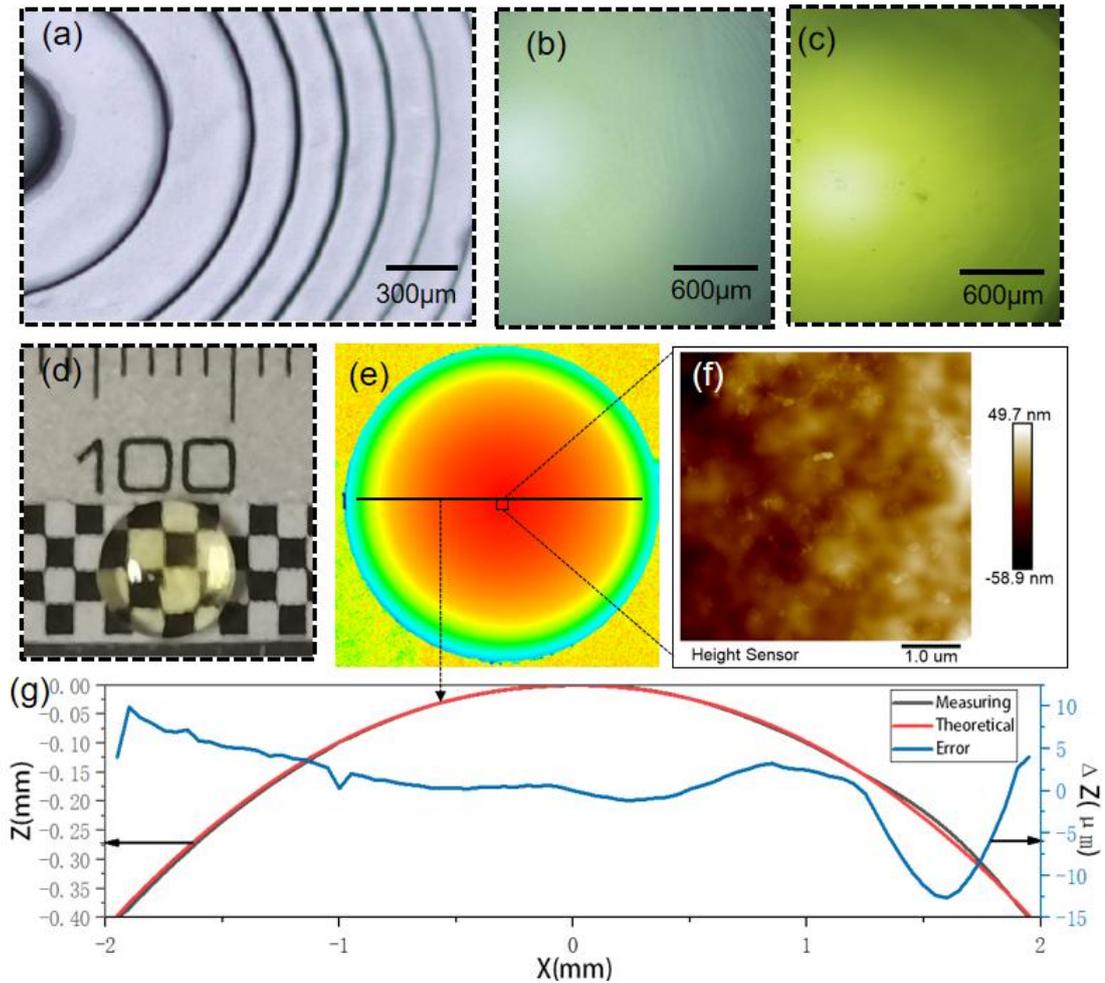

Figure 5 (a) Printed local surface of the lens with steps (b) Local surface of the lens obtained after immersion (c) Smooth local surface of the lens obtained after secondary curing (d) Lens size (e) Surface morphology of the lens obtained through confocal microscopy. (f) The surface roughness (g) obtained by atomic force microscopy within a square in lens (e) compares the actual contour of the lens with the design contour.

Figure 5e shows the surface morphology obtained by optical white light interferometry. Atomic force microscopy was used to test the surface roughness of the lens within the area range of 5 μm x 5 μm in Figure 5e, and the measurement results showed a roughness of 8.18nm. Figure 5g shows the comparison between the experimentally measured surface profile of the 3D printed lens and the original design to characterize manufacturing accuracy and accuracy, and provides the error between the two. It can be seen that the experimentally measured surface profile matches the design values, but in some positions, there may be some deviation due to improper resin filling.

The optical imaging resolution of the lens structure was evaluated by the United States Air Force (USAF) 1951 standard resolution test target. The image captured through the lens was taken in transmission mode using an optical microscope with a 10 x objective lens and a CCD camera, using green light as the light source. Figure 6b shows the theoretical values of MTF for three different fields of view with Y angles of 0°, 5°, and 10°, respectively. The target image captured by CCD without adding a lens is shown in Figure 6c. Figure 6d shows the effect after inserting a lens with steps. It can be seen that the uneven lens also has a certain magnification effect, but the imaging quality is not good. Figure 6e shows an image with a magnification of approximately 1.1 obtained by placing a lens on a smooth surface, and we can clearly observe all elements in groups 4-5. Compared with the original test target pattern that was not imaged through the lens structure, the imaging pattern can still be clearly distinguished through the lens structure. Therefore, the 3D printed lens structure exhibits significant optical quality in the visible spectrum. Figure 6f shows the intensity distribution of the three elements in the enlarged sixth group. For quantitative analysis, we calculated the intensity contrast intensity distribution of elements 1-3 in the sixth group along the horizontal direction, with contrasts of approximately 52%, 40%, and 38%, respectively. Figure 8f shows that 38% corresponds to element 3, with a resolution of 80lp/mm. The results indicate that 80lp/mm is in good agreement with the actual experimental results.

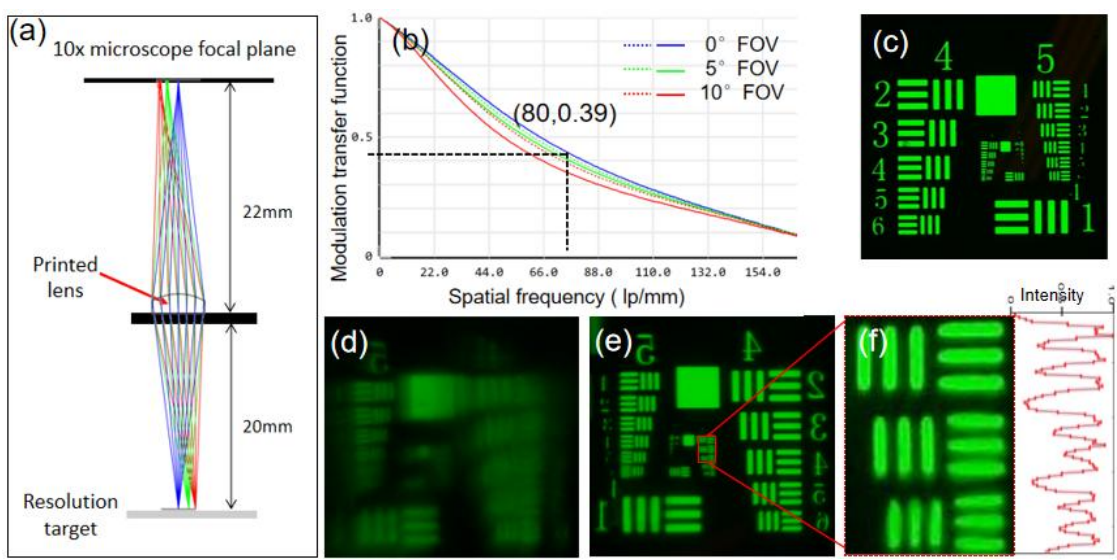

Figure 6: Imaging of resolution targets (a) Resolution testing of 3D printed lens structure. The illustration shows the corresponding test positions on the 3D printed lens structure. (b) MTF corresponds to the theoretical value of the printed lens.(c) The target image without adding a lens (d) is the effect image with a stepped lens (e) is the effect image with a smooth lens (f) is the magnified image of the 6th group of elements 1-3. The brightness contrast intensity distribution of the 6th group of elements 1-3 in the imaging image.

### 3.3 Application of Non spherical Lenses

In order to quickly demonstrate actual imaging, the produced aspherical lens was directly attached to the rear camera of the smartphone (Figure 7a). Figures 7b and c show close-up views of the photon sieve captured by the smartphone at the same distance without and with the lens added. When not using a lens, due to the small object distance, the imaging is behind the CCD of the phone, and the focus is not accurate, which will make the discharged photos unclear; When placing the printed lens in front of the phone lens, still place the photon sieve in the same position as before. Unlike just now, due to the convergence effect of the lens, the light falls onto the CCD, resulting in a clear image. We also captured local details of the leaves of the Blood Activating Pill, as shown in Figure 7d. We can clearly see the veins and surface fuzz of the leaves. The test results confirm that 3D printing lenses can be well applied to commercial cameras to provide high-performance images.

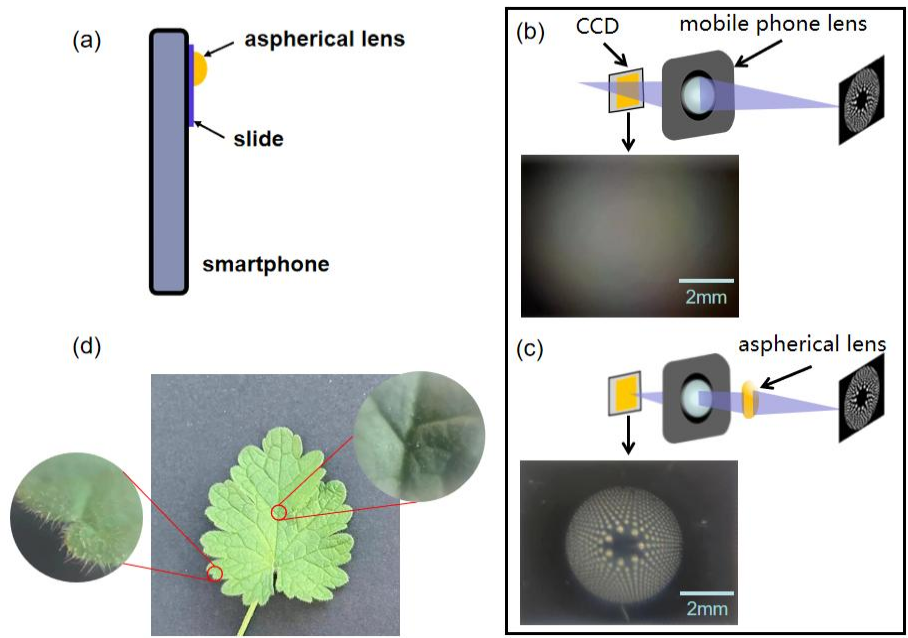

Figure 7 (a) Side view of sticking the lens to the mobile phone camera (b) Photon screen image when the lens is not placed behind the CCD (c) Clear image of the photon screen when the lens is placed on the CCD (d) Local display details of the leaves of the Blood Activating Pill are captured using a mobile phone with the lens attached

## 4 Conclusion

This article uses digital light processing technology to print an even order aspherical lens with a height of 2mm and a diameter of 5mm within 38 minutes. During this process, the influence of material ratio and printing parameters on the accuracy of printed parts was explored. Using fluoride solution as the isolation layer and using vertical pull-up immersion method to eliminate steps, a lens with good imaging quality was finally obtained. The optical performance of the smooth lens was tested, and its resolution was characterized as 80lp/mm and surface roughness was 8.18nm. This proves the feasibility of the method proposed in this article, and also implies that it has great potential in the production of optical components. And digital light processing technology, this flexible and efficient printing method, also has huge application prospects.